\newdimen\figwidth
\def\a{\alpha}
\def\b{\beta}
\def\G{\Gamma}
\def\A{{\cal A}}
\def\K{{\cal K}}
\def\CP{{\Bbb P}}

\documentstyle[epsf]{article}	\def\Bbb{\bf}	\def\ltimes{\times}
\begin{document}
\begin{titlepage}
\null \vskip -0.6cm
\parbox{4cm}{PAR--LPTHE 94--08}
\hfill  \date{February 94} \hfill
\parbox{5cm}{hepth@xxx/9403130}
\vskip 1.4truecm
{
\begin{center}
BAXTERIZATION,  DYNAMICAL SYSTEMS,\\
AND THE SYMMETRIES OF INTEGRABILITY \\
\vskip 1truecm
C.-M. Viallet
\footnote{
\obeylines
Laboratoire de Physique Th\'eorique et des Hautes Energies, Centre National de
la Recherche Scientifique,
Universit\'e de Paris 6- Paris 7, Tour 16, $1^{\rm er}$ \'etage, bo\^\i te 126.
4 Place Jussieu/ F--75252 PARIS Cedex 05 / FRANCE / e-mail:
viallet@lpthe.jussieu.fr }

\end{center}
}

{\noindent {\bf Abstract}.
We resolve the `baxterization' problem with the help of the automorphism group
of the Yang-Baxter (resp. star-triangle, tetrahedron, \dots) equations.
This  infinite group of symmetries  is realized as a non-linear (birational)
Coxeter group  acting on matrices, and exists as such,  {\em  beyond the narrow
context of strict integrability}.  It yields among other things  an unexpected
elliptic parametrization of the non-integrable sixteen-vertex model.
It provides us with  a class of discrete dynamical systems, and we address some
related problems, such as   characterizing  the complexity of iterations.}

\vfill
\begin{center}
To appear in the proceedings of the  the \\
Third Baltic Rim Student Seminar  HELSINKI (September 1993) \\
\vskip 1truecm
work supported by CNRS \\
\end{center}
\end{titlepage}

\section{Introduction}

The results presented here originate in  a long-standing collaboration of
M.~Bellon, J.M.~Maillard and the
author~\cite{BeMaVi91a,BeMaVi91b,BeMaVi91c,BeMaVi91d,BeMaVi91e,BeMaVi92,BeBoMaVi93}, augmented by  further elaborations (as~\cite{BeMaRoVi92,FaVi93}).

The Yang-Baxter equations and their variations (star-triangle equations,
tetrahedron equations,...) are nowadays considered as a characteristic  feature
of integrability  of mechanical systems (classical hamiltonian systems, as well
as quantum and statistical
systems)~\cite{On44,Gu67,Ba81,Ya67,TaFa79,FaSkTa80,Sk80,KuSk82,Fa82,Ga83,Sk92}\dots.
This field has expanded very much for the last twenty years.
We shall not dwell here on its innumerable developments, and relegate general
considerations to the end.

 We shall start from  the Yang-Baxter equations and view them as an algebraic
system of equations on matrix entries.

We first recall (section  \ref{ybev}) what  the two most common forms of the
equations for vertex models are.

In section \ref{gammavertex2} we introduce certain groups generated by
involutions (Coxeter groups) \cite{CoMo65}, together with some realizations in
terms of birational transformations of projective spaces. These realizations
are generically denoted $\G$ in the sequel.

In section \ref{autoyb} we build up,  from the groups $\G$,  an infinite  group
of automorphisms  (denoted $\A$) of the Yang-Baxter equations.
One should keep in mind that  the equations we analyze  form an {\em
overdetermined} system, making the  existence of a large group of symmetry
quite remarkable.

We next show, for the paradigmatic example of the Baxter symmetric eight-vertex
model and   with both a picture and algebraic results, how our automorphism
group reconstructs the well known  elliptic curves of solutions. This is a
first solution to the `baxterization' problem~\cite{Jo90}.

In section \ref{other} we describe more groups of matrix transformations. These
are very similar to the  ones of section \ref{gammavertex2}, but are adapted to
the star-triangle relations,  tetrahedron equations, and higher dimensional
generalizations.

In section \ref{baxterization_full} we complete the solution to the
baxterization problem, emphasizing the notions of $\G$-covariant versus
$\G$-invariant quantities.

We next produce (section~\ref{invariants}) an algorithm to calculate the
covariant polynomials as well as the algebraic  invariants.

One of the features of $\G$ is that it is defined outside of the space of
solutions of the  Yang-Baxter equations,   and we may look at its action {\em
beyond integrability}. It becomes  an ordinary discrete dynamical system on the
space of parameters, possibly non-conservative.
In section \ref{seize}, we  describe this action for the general 16-vertex
model, i.e. the most general two-state vertex model on a two-dimensional square
lattice, which is known {\em not to be integrable}.
The outcome is quite surprising: the orbits of  $\G$ stay within 1-dimensional
subvarieties of the 15-dimensional space of parameters. This reveals  the
actuality of an amazingly large number of algebraically independent invariants
of $\G$, and yields  an elliptic parametrization of the non-integrable model,
without  reference to the Yang-Baxter equations.

In section \ref{complexity} we turn to the notion of complexity of the
iterations and show how it fits with the existence of algebraic invariants.

In the last section (\ref{classique}), we exemplify some of our results on a
definite system, related to a three dimensional vertex model.

\section{The Yang-Baxter equations  (vertex models)}
\label{ybev}
Many   presentations of the Yang-Baxter equations appear in the  literature,
dividing into two classes:

The first  class contains a parameter, also called spectral
parameter~\cite{Fa82},  and reads, for vertex models:
\begin{eqnarray}
  \label{yb}
     &   \sum_{\alpha_1,\alpha_2,\alpha_3} R^{i_1i_2}_{\a_1\a_2}
(\lambda_1,\lambda_2)
                R^{\a_1i_3}_{j_1\a_3} (\lambda_1,\lambda_3)
R^{\a_2\a_3}_{j_2j_3} (\lambda_2,\lambda_3) \qquad \qquad & \\
   & \qquad \qquad  =\sum_{\b_1,\b_2,\b_3} R^{i_2i_3}_{\b_2\b_3}
(\lambda_2,\lambda_3)
          R^{i_1\b_3}_{\b_1j_3} (\lambda_1,\lambda_3) R^{\b_1\b_2}_{j_1j_2}
(\lambda_1,\lambda_2).& \nonumber
\end{eqnarray}

In this system of equations, $R$ is a matrix of size $q^2\times q^2$,whose
entries are functions of the parameters $\lambda_i$, usually through the
difference $\lambda_i - \lambda_j$ (at least when $\lambda$ labels a point on
an elliptic curve).

The parameter first appeared as parametrizing the solution. It was later
understood as a ``spectral parameter'' in the context of the  quantum inverse
scattering method. The origin of the parameter is one of the issues we wish to
address.

A given family of models is obtained when one specifies the size $q$ and  the
relations between the entries of $R$.
Typical relations are equalities between different entries or vanishing of some
others (see for example section~\ref{baxterization}).

Since the entries of $R$ are allowed to  be different for the three copies of
$R$ entering (\ref{yb}), one may rewrite~(\ref{yb}) with three explicitly
different  matrices $A, B$, and $C$   of the same size $q^2\times q^2$.
\begin{equation}
 \sum_{\alpha_1,\alpha_2,\alpha_3} A^{i_1i_2}_{\a_1\a_2}
                B^{\a_1i_3}_{j_1\a_3}  C^{\a_2\a_3}_{j_2j_3}
     =\sum_{\b_1,\b_2,\b_3} C^{i_2i_3}_{\b_2\b_3}
          B^{i_1\b_3}_{\b_1j_3}  A^{\b_1\b_2}_{j_1j_2}
\end{equation}
or shortly
\begin{equation} \label{abc}
  A_{12} B_{13} C_{23} = C_{23} B_{13} A_{12}
\end{equation}
with now usual notations.
The latter form has the interest of not referring to any explicit
parametrization.

The second class contains no parameter, and is sometimes called ``constant''
Yang-Baxter equation. It reads:
\begin{equation} \label{cyb}
  R_{12} R_{13} R_{23} =R_{23} R_{13} R_{12}
\end{equation}
that is to say like~(\ref{abc}) but with  $A=B=C=R$.

To go from equation (\ref{yb}) to equation (\ref{cyb}) requires essentially  an
adequate choice of  $\lambda_1, \lambda_2,  \lambda_3$, namely $\lambda_1 =
\lambda_2 =  \lambda_3$.
The reverse move, i.e. to recover the parameter dependence from a constant
solution was given the name of `baxterization'~\cite{Jo90}.

The  {\em use}  of the spectral parameter  is clear in the context of the Bethe
ansatz, since it serves building up the generating functional for {commuting
conserved quantities}, but we shall not be concerned with this aspect. Its {\em
origin} will be explained in the next sections.

\section{Some operations on matrices: the groups $\G$}
\label{gammavertex2}

We describe here some elementary  operations on matrices of various sizes.
The matrices we consider are defined up to an overall multiplicative factor.
The space of parameters  is thus some projective space $\CP_n$  with the
entries of the matrix as  homogeneous coordinates.

Let us start with the matrix $R$ of a two-dimensional $q$-state vertex model on
a square lattice.
$R$ has  the structure of a tensor product and the indices of $R$   are  pairs
of indices:
\begin{equation}
R^{ij}_{kl} \qquad i,j,k,l=1..q
\end{equation}
We may define  the inverse $I$ up to a factor (well defined in the projective
space)
\begin{equation}
\sum_{\a\b} (IR)^{ij}_{\a\b} R^{\a\b}_{kl}=\mu \; \delta^{i}_{k} \delta^j_l
\qquad i,j,k,l=1..q
\end{equation}
with $\mu$ an arbitrary  multiplicative factor, the transposition $t$:
\begin{equation}
(tR)^{ij}_{kl}= R^{kl}_{ij} \qquad i,j,k,l=1..q
\end{equation}
as well as  two {\em partial transpositions} $t_l$ (index $l$ for left) and
$t_r$ respectively by:
\begin{equation}
(t_lR)^{ij}_{kl}= R^{kj}_{il} \qquad (t_rR)^{ij}_{kl}= R^{il}_{kj} \qquad
i,j,k,l=1..q
\end{equation}
Of course
\begin{equation}
t=t_l\; t_r= t_r\; t_l, \qquad I^2=t^2=t_l^2=t_r^2=1,
\quad \mbox{ and } \quad I\; t = t \; I
\end{equation}
However
\begin{equation}
t_l\; I \neq I\; t_l \quad \mbox{ and } t_r\; I \neq I\; t_r \nonumber
\end{equation}
The two partial transpositions {\em do not commute with the inversion while
their product  $t$ does}.
The transformation $t_l I$ and  $t_r I$ {\em are of infinite order}.
Notice of course that they may have finite orbits when acting on certain
non-generic matrices.

The group $\G$ generated by $t_l, t_r$ and $I$ essentially consists  of the
iterates of $t_l I$ and its inverse $It_l$  (or $t_r I$ and $It_r$) up to
multiplication by an element of its center. It is generated by involutions
realized as rational (possibly linear) transformations of some projective space
$\CP_n$.

We shall describe in  section~\ref{other} more instances of groups $\G$ acting
as rational  transformations of matrices, and generated by involutions.

\section{The automorphisms of the Yang-Baxter equations} \label{autoyb}

Suppose we have a solution $(A,B,C)$ of~(\ref{abc}):
\begin{equation} \label{abcloc}
  A_{12} B_{13} C_{23} = C_{23} B_{13} A_{12}
\end{equation}
We may take the partial transpose $t_1$ of (\ref{abcloc})  along space 1 and
get:
\begin{equation} \label{t1yb}
(t_1B_{13}) \; (t_1A_{12}) \; C_{23} = C_{23} \; (t_1A_{12}) \;(t_1B_{13})
\end{equation}
Taking the partial transpose $t_2$ of~(\ref{t1yb}) yields
\begin{equation}
(t_1B_{13}) \; (t_2C_{23}) \; (t_1t_2A_{12}) = (t_1t_2A_{12}) \; (t_2C_{23}) \;
(t_1B_{13})
\end{equation}
that is to say:
\begin{equation} \label{t1t2yb}
(t_lB)_{13}\; (t_lC)_{23}\; (tA)_{12} = (tA)_{12}\; (t_lC)_{23} \; (t_lB)_{13}
\end{equation}
Multiplying both sides of~(\ref{t1t2yb}) to the left and the right by the
inverse $ItA$ of $tA$ gives:
\begin{equation}
(tIA)_{12} \; (t_lB)_{13} \; (t_lC)_{23} =  (t_lC)_{23} \; (t_lB)_{13}
\;(tIA)_{12}
\end{equation}
which shows that the transformation
\begin{equation}
\K_A: (A,B,C) \longrightarrow (tIA, t_lB, t_lC)
\end{equation}
takes a solution of~(\ref{abc})  into a solution of~(\ref{abc}).

We could similarly construct $\K_B$ (resp. $\K_C$) which inverts and transposes
$B$ (resp. $C$) and acts by partial transpositions on $A$ and $C$ (resp. $A$
and $B$).

The automorphisms $\K_A, \K_B, \K_C$ are involutive and generate an infinite
group of automorphisms $\A$ of the Yang-Baxter equations. It is possible to
show that  $\A$ is isomorphic to the Weyl group of an affine Lie algebra of
type $A_2^{(1)}$~\cite{BeMaVi91c,BeMaVi91d}.

The group $\G$ appears when  we look at  the action of $\A$ on one of the
individual copies  of $R$ entering  equation~(\ref{yb}).
However, the  action of $\G$ is defined without reference to~(\ref{yb}), and we
will now concentrate on it, and somehow forget about~(\ref{yb}).

\section{Baxterization of the Baxter model} \label{baxterization}

The exemplary  two-dimensional integrable vertex  model is the Baxter
model~\cite{Ba71,Ba71b,Ba81}, with matrix of Boltzmann weights:
\begin{equation} \label{baxmat}
	R=\pmatrix{	a&	0&	0&	d \cr
			0&	b&	c&	0 \cr
			0&	c&	b&	0 \cr
			d&	0&	0&	a \cr}
\end{equation}
where one sees the vanishing conditions and equality relations between the
entries.
It is straightforward to see that the form (\ref{baxmat}) is stable by all the
generators of $\G$. We say that the pattern (\ref{baxmat}), i.e. the collection
of vanishing conditions and equalities between entries is admissible
(see~\cite{BeMaVi91a,BeMaVi91b}).

The transformations  $t_l$ and $t_r$ read:
$$ a \rightarrow a, \quad b \rightarrow b, \quad c\rightarrow d, \quad
d\rightarrow c,$$
and  $I$ reads
\begin{eqnarray*}
        a \rightarrow 		{a\over a^2-d^2}, \quad
                b \rightarrow {b\over b^2-c^2}, \quad
        c \rightarrow	 	{-c\over b^2-c^2}, \quad
                d \rightarrow {-d\over a^2-d^2}.
\end{eqnarray*}

One  may look at the action of $\G$ on $R$. Take one starting point in the
space $\CP_3$ of parameters, iterate  $It_l$ on it and just look at the orbit.
One  gets the following picture, showing a three-dimensional perspective of the
orbit, in inhomogeneous coordinates $u=b/a$, $v=c/a$, $w=d/a$, with starting
point $(*)$:

\par
\epsfxsize=\figwidth
\centerline { \epsffile[0 0 512 350]{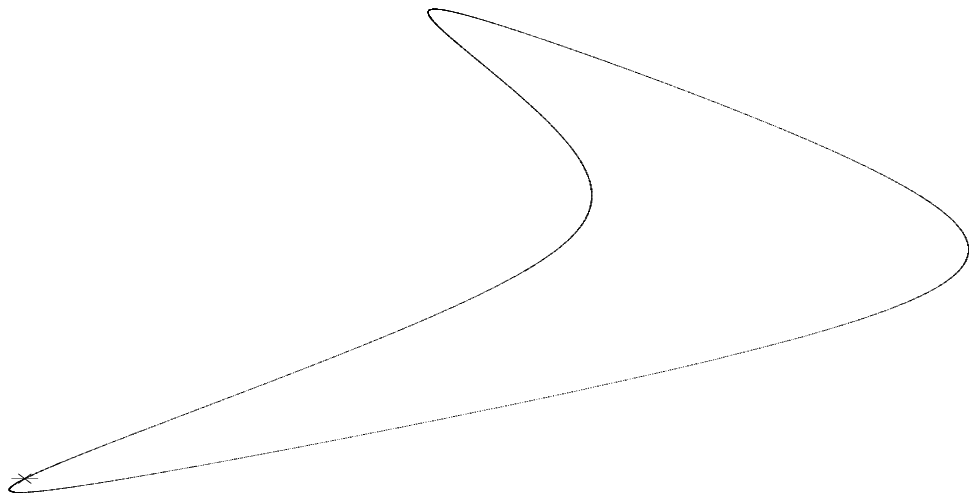} }
\centerline{\it  Figure 1: `Baxterization' of the Baxter model: a perspective
view}
\par
\smallskip

One sees from this figure that the orbit lies inside a curve. The action of
$\G$ must have at least two algebraically independent invariants.
Indeed there are two such invariants:
\begin{eqnarray}
        \Delta_1 = {a^2 + b^2 - c^2 - d^2 \over ab+cd },    \qquad
        \Delta_2 = { ab - cd \over ab + cd }.
\end{eqnarray}
The curves $\Delta_1=constant, \Delta_2=constant$, are precisely the elliptic
curves which appear in the solution of~(\ref{yb}), {\em but we have obtained
them with the form~(\ref{baxmat}) as only input, and in particular without
having to solve~(\ref{yb})}. The resolution of~(\ref{yb}) becomes
straightforward once the adequate parametrization of the above curves is used.
The orbit we get, although discrete, tends to be   dense inside one dimensional
curves. The phenomenon is similar to the iteration of a rotation (obtained as
the composition of two axial  symmetries) with irrational rotation number.

Warning: It is in no way the existence of these curves  which ensures the
compatibility of the Yang-Baxter equations!

\section{Other transformation groups} \label{other}

There are many other equations like~(\ref{yb}). They also have automorphism
groups generated by (bi)rational transformations. This means we may construct
many different realizations  $\G$. Here are a few more.

\subsection{From spin model with interactions along the edges}
\label{fromspin}
As was shown in~\cite{BeMaVi91c}, the  star-triangle equation also has an
infinite group of automorphisms. The matrices under consideration are ordinary
$q\times q$ matrices $m$, of entries $m_{ij}$.
The generators of $\G$ are  the matrix inverse $I$ as previously
\begin{equation} \label{spini}
\sum_{\a} (Im)_{i\a} \; m_{\a j} = \mu \; \delta_{ij}
\end{equation}
with $\mu$ an arbitrary multiplicative constant, and the element-by-element
inverse (Hadamard inverse):
\begin{equation} \label{spinj}
J: \qquad m_{ij} \longrightarrow 1/m_{ij}
\end{equation}
$I$ and $J$ are two non-commuting rational involutions. The product
$\varphi=IJ$ is a birational transformation of  infinite order.

\subsection{From vertex models in three  (and more) dimensions}
\label{tetraedre}
The matrices under consideration in this case again  have  multi-indices. The
matrices are of size  $q^3\times q^3$, with  entries of the form:
\begin{equation}
 R^{ijk}_{lmn}
\end{equation}
By examining the symmetries~\cite{BeMaVi91d,BeMaVi91e} of the tetrahedron
equations~\cite{Za81,JaMa82b,MaNi89}:
\begin{equation}
        R_{123} R_{145} R_{246} R_{356} =  R_{356}  R_{246}  R_{145}  R_{123}.
        \label{tetra}
\end{equation}
with the usual notations, one is lead to the group $\G$ generated by the
inverse $I$ and the {\em three} partial transpositions $t_g$, $t_m$ and $t_d$
with
\begin{equation}
        (t_gR)^{i_gi_mi_d}_{j_gj_mj_d} = R^{j_gi_mi_d}_{i_gj_mj_d},
\end{equation}
and similar definitions for $t_m$ and $t_d$.

One could as well consider multi-index matrices of size $q^d \times q^d$
written in the form $M^{i_1 i_2 \dots i_d}_{j_1 j_2 \dots j_d}$.
Section \ref{autoyb} deals with $d=2$. The tetrahedron equation corresponds to
$d=3$, and so on.
There exist  $d$ different partial transpositions $t_1,\, t_2, \dots, t_d$ with
obvious definitions. Combined with the inverse $I$, they generate an {\em
infinite\/} group $\G$. We will return to the question of the size of these
groups in sections~\ref{invariants} and~\ref{complexity}.

\subsection{More groups}
One  may further enrich the representations by imposing constraints on the
entries of the matrices, provided the transformations are compatible with these
constraints (see~\cite{BeMaVi91a} for the notion of admissible patterns).
This yields realizations on projective spaces of lower dimensions.

If the dimension  $n$ of the projective space is 1, the only birational
transformations are homographies.
If $n=2$, the group, called Cremona group~\cite{De71}, is already much larger.
Its elements may however be written as products of linear transformations and
the basic blow-up $J$.
The interesting subgroup  made out of polynomial and polynomially invertible
transformations of the plane has a more elementary
structure~\cite{Ju42,FrMi89}.
When $n$ is larger than 2, the structure of the group of birational
transformations is much more complex.

Let us  stress that, at the level of the  realization, there may exist
additional relations between the generators, possibly making it  finite.

\section{Baxterization}
\label{baxterization_full}

It appears clearly in  sections \ref{autoyb} and  \ref{baxterization} that, for
the Baxter model, the existence of the parameter comes from the existence of
sufficiently many algebraic invariants of $\G$ to confine its orbits, and
therefore  the ones of $\A$, to curves.
The general situation is more subtle, but it always amounts to the following:
find the  $\G$-covariant varieties passing through a constant solution, and/or
look for solutions only on covariant varieties.

Different situations may appear, and we illustrate them by specific examples:

-- the simplest one was encountered in section  \ref{baxterization}, recalling
that the only input information was the form of the matrix (\ref{baxmat}). The
parameter is just the uniformizing parameter for the elliptic curve which pops
out of the action of $\G$ on any of its generic points. In this case, the
action of $\G$ {\em has  INVARIANTS to imprison the orbit}, and the entire
space of parameters is foliated by invariant subvarieties (curves for the
Baxter model).
Each generic spectral curve has an infinite group of automorphisms, and they
consequently all have genus 0 or 1. The solution of (\ref{abc}) is obtained in
the following way: choose any point $A$ in the space of parameters. Through $A$
passes one curve of the foliation. Choose $B$ anywhere on this curve. The last
matrix $C$ is completely determined then. What is special about the Baxter
model is that there is no restriction on the first choice ($A$), within the
space $\CP_3$ of parameters. This is far from being the general occurrence.

-- a more subtle situation is encountered for the chiral Potts model (as well
as for  the free fermion condition~\cite{FaWu70,Fe73,Kr81}). Integrability
appears on a algebraic variety of which all points have a periodic orbit under
$\G$. This variety is automatically globally invariant by $\G$, but there may
very well be NO INVARIANT of the action of $\G$. Besides,  the resolution of
the Yang-Baxter equations may lead to consider further algebraic conditions.
Solutions are to be found on a possibly  isolated subvariety of the space of
parameters. In this situation, the spectral curve may have  genus higher than 0
or 1, since it has only a finite number of
automorphisms~\cite{AuCoPeTaYa87,BaPeAu88,HaMa88}.

-- a similar situation is encountered in the case of the Jaeger-Higman-Sims
model~\cite{Ja91,Ha94}.
The parameter space of the model is $\CP_2$. The matrix of Boltzmann weights
belongs to a 3-dimensional abelian algebra of matrices of size $100\times 100$,
with three generators ${\bf 1}, A, {\cal J}-{\bf 1}-A$ where ${\bf 1}$ is the
unit matrix,  $A$  the adjacency matrix of the Higman-Sims graph, and ${\cal
J}$ the  matrix with all entries equal to $1$.
Two products exist in this algebra, respectively the usual matrix product and
the  element by element product. The symmetry group is generated by the
inverses $I$ and $J$, corresponding to these two products, as in  section
(\ref{fromspin}), equations (\ref{spini}, \ref{spinj}).
The two inverses are related by a collineation $C$ (linear map of $\CP_2$):
$$ I = C^{-1} J C $$
In terms of homogeneous coordinates $[x_0,x_1,x_2]$:
\begin{eqnarray*}
& J: [x_0,x_1,x_2]  \rightarrow [x_1 x_2, x_0 x_2, x_1 x_2] & \\
& C = \left [\begin {array}{ccc} 1&22&77\\1&-8&7\\1&2&-3\end {array}\right ]
 \label{jaeger} &
\end{eqnarray*}
We can find three  $\G$-invariant subvarieties of $\CP_2$: one line D, and two
hyperbolae H1, H2:
\begin{eqnarray*}
& D: & {\it x_2}-{\it x_1} =  0 \\
& H1:&  {\it x_2}^{2}+3\,{\it x_1}\,{\it x_2}-3\,{\it x_0}\,{\it x_2}-{\it
x_0}\,{\it x_1}=0 \\
& H2: & {\it x_1}\,{\it x_2}+2\,{\it x_1}^{2}-2\,{\it x_0}\,{\it x_2}-{\it
x_0}\,{\it x_1} =0
\end{eqnarray*}
Hyperbola  $H1$ is the one of reference~\cite{Jo89}.
The two hyperbolae contain an infinite number of singular points, and we know
this prevents the existence of an invariant~\cite{FaVi93}, but not of covariant
polynomials, and a fortiori not of covariant ideals.

\section{Covariants and  invariants}  \label{invariants}

It is fortunately possible to go further in the analysis of the invariant
subvarieties~\cite{FaVi93}.

We start from a group $G$ generated by  by $\nu$  involutions $I_1, I_2, \dots,
I_k, \; (k=1\dots \nu)$, {\em verifying no relations other than the involution
property}.
The group $G$ is infinite and there are two essentially different situations.

If $\nu=2$, the group is the infinite dihedral group ${\bf Z}_2 \ltimes {\bf
Z}$, and all elements may uniquely be written $I_1^\alpha(I_1I_2)^q$, with
$\alpha=0,1$ and $q\in {\bf Z}$. The number of elements of given length $l$ is
2.

If $\nu \geq 3$, the number of elements of length $l$ grows exponentially with
$l$, and the group is in a sense bigger (still countable).

As an example,  for the groups described in section~\ref{tetraedre},  the
number $\nu$ of generators depends on the dimension $d$ of the lattice: it is
just $2^{d-1}$ so that if $d=2$,  $G$ is generated by two involutions  and   if
$d\ge 3$, $G$ is generated by more than three involutions.

We then construct  various  realizations $\Gamma$ of $G$ by explicit
transformations of some projective space.
They are  obtained by specifying the realization of the generators.
We  use the same  notation $I_k$ for the generators of $G$ and their
representatives in $\G$.
The realizations $\G$ of $G$ may be written as polynomial transformations in
terms of the homogeneous coordinates.

Each involution $I_k$ defines a  characteristic polynomial $\phi_k$ of degree
$d_k^2-1$  in the following manner. The $I_k$ being involutions,  $I_k^2$
appears as the multiplication by a degree $d_k^2-1$ polynomial
$\phi_k(x_0,\ldots,x_n)$.

A $\G$-covariant polynomial $P$ verifies:
\begin{equation}
\label{cov}
P(\gamma ( x)) = a( \gamma, x) P(x) \qquad \forall \gamma \in \G, \forall x \in
\CP_n
\end{equation}
The coefficient  $ a( \gamma, x)$ has to fulfill the cocycle condition:
\begin{equation} \label{cocycle}
a(\gamma_1 \gamma_2,x) = a(\gamma_1, \gamma_2\; x)a(\gamma_2,x)
\end{equation}
Indeed  relation (\ref{cov}) demands
\begin{eqnarray*}
P(\gamma_1 \gamma_2 x) 	&=& a(\gamma_1 \gamma_2,x) P(x) \\
			&=& a(\gamma_1,  \gamma_2 \; x) P(\gamma_2 x) \\
			&=& a(\gamma_1,  \gamma_2 \; x) a(\gamma_2,x)P(x)
\end{eqnarray*}

The cocycle $a$ will be completely determined by the values of $a(I_k,x),
k=1\dots \nu$. These values may be  found easily:
when applied to $\gamma_1=\gamma_2=I_k$, condition (\ref{cocycle}) shows that
$a(I_k,x) $ has to divide a suitable power of $\phi_k$.

Finding an invariant $\Delta=P/Q$ implies  finding two polynomials $P$ and $Q$
of the same degree,   which transform the same way (are covariant with the same
cocycle) under all the generators, i.e:
\begin{equation} \label{linsys}
P(I_k(x)) =  a(I_k,x) \cdot P(x) \qquad \mbox{and} \qquad  Q(I_k(x)) =
a(I_k,x) \cdot Q(x)
\end{equation}

Once the cocycle $a$ is chosen, solving (\ref{linsys}) becomes a handable
linear problem, of which the compatibility can be further
studied~\cite{FaVi93}.
We have proved that  the proliferation of singularities impeaches the existence
of any invariants, but the converse is not true.
One of the  outcomes is that, in this general setting,  the existence of
invariants is exceptional.

\bigskip
Finding all invariant subvarieties is  more tricky.
What we have obtained so far is  invariant subvarieties {\em of codimension 1},
or subvarieties determined by invariants of $\G$.
Smaller subvarieties have more than one equation, and   (\ref{cov}) has to be
replaced by a matrix relation:
\begin{equation}
\label{matcov}
\Pi(\gamma ( x)) = A( \gamma, x) \Pi(x) \qquad \forall \gamma \in \G, \forall x
\in \CP_n
\end{equation}
where $\Pi$ is a vector constructed from the (non canonical) list of equations,
and $A$ a matrix. In other words we demand that the ideal defining the
subvariety be invariant by $\G$, keeping in mind that the subvariety we look
for does not have to be a complete intersection.

Let us describe an example, coming from the hard hexagon
model~\cite{BaPe82,AvMaTaVi90}. Consider the two involutions of $\CP_4$ given
in terms of homogeneous coordinates $[{\it x_0}, {\it x_1},{\it x_2},{\it
x_3},{\it x_4}]$:
\begin{eqnarray*}
i_1& \longrightarrow & [{\it x_5}\,{\it x_1}\,{\it x_4},{\it x_4}\,\left ({\it
x_0}\,{\it x_5}-{\it x_2}^{2}\right ),-{\it x_2}\,{\it x_1}\,{\it x_4},{\it
x_1}\,\left ({\it x_0}\,{\it x_5}-{\it x_2}^{2}\right ),{\it x_0}\,{\it
x_1}\,{\it x_4}] \\
i_2 & \longrightarrow & [{\it x_4}\,{\it x_2}\,{\it x_5},-{\it x_1}\,{\it
x_2}\,{\it x_5},{\it x_5}\,\left ({\it x_0}\,{\it x_4}-{\it x_1}^{2}\right
),{\it x_0}\,{\it x_2}\,{\it x_5},{\it x_2}\,\left ({\it x_0}\,{\it x_4}-{\it
x_1}^{2}\right )]
\end{eqnarray*}
The two inverses are related by a linear transformation, i.e  $i_2 = \tau \;
i_1 \tau$ with
\begin{equation} \nonumber
\tau: [{\it x_0},{\it x_1},{\it x_2},{\it x_4},{\it x_5}] \longrightarrow[{\it
x_0},{\it x_2},{\it x_1},{\it x_5},{\it x_4}]
\end{equation}
and
$\phi_1= i_1^2 = {\it x_1}^{2}{\it x_4}^{2}\left ({\it x_0}\,{\it x_5}-{\it
x_2}^{2}\right )^{2}$.
If \begin{eqnarray*}
& \Pi_1 =
{x_0}^{2}{x_2}^{2}{x_4}+{x_0}^{2}{x_1}^{2}{x_5}-{x_0}^{3}{x_4}\,{x_5}-{x_4}^{2}{x_5}\,{x_2}^{2}-{x_4}\,{x_5}^{2}{x_1}^{2}+{x_0}\,{x_4}^{2}{x_5}^{2}-{x_1}^{2}{x_2}^{2}{x_0} &\\
& \Pi_2 = {x_0}^{2}-{x_4}\,{x_5}-\lambda \; {x_1}\,{x_2} &
\end{eqnarray*}
with $\lambda$ a free parameter, then we do have relation (\ref{matcov}) with
a matrix $$ A_1 = \left [\begin {array}{cc} {\it x_1}^{3}{\it x_4}^{3}\left
({\it x_0}\,{\it x_5}-{\it x_2}^{2}\right )^{2}&0\\-{\it x_4}&-{\it
x_4}^{2}\left ({\it x_0}\,{\it x_5}-{\it x_2}^{2}\right )\end {array}\right ]
$$
and
$$ A_1(i_1(x)) A_1(x) = \pmatrix {	\phi_1^5(x) & 0 \cr
					0  & \phi_1 ^2(x) \cr } $$
Both $\Pi_1$ and $\Pi_2$ are unchanged by $\tau$, $\Pi_1$ is covariant, and
consequently the variety $\Pi_1=0$ is invariant by $\G$. In contrast, $\Pi_2$
is not covariant and only the intersection of $\Pi_2=0$ with $\Pi_1=0$ is
invariant by $\G$.

\section{The sixteen-vertex model}
\label{seize}

We abandon here the strict context of integrability, and turn to  problems of
{\em discrete dynamical systems, defined in the space of parameters of the
models}.

The field of discrete dynamical system is a well developed one. The notion of
integrable map goes back, in the context of hamiltonian dynamics,  to
Poincar\'e, who founded the subject (\cite{Po52,Po92}, see
also~\cite{Bi17,Wi81}).  The last 30 years have seen the subject expand a lot,
specifically  with the advent of computer calculations.
In order to gain  simplicity, one was   lead long ago  to  studying iterations
of  polynomial and polynomially invertible transformations of the
plane~\cite{He76}), eventually renouncing to hamiltonian structures.
Analytic maps have also been fruitfully analyzed, especially in one complex
dimension~\cite{Si42,Yo84}.
More recently, some remarkable multi-parameter families of maps, eventually
integrable, have  been constructed  from soliton
equations~\cite{QuRoTh88,QuRoTh89,MoVe91,GrRaPa91,PaNiCa90,Ra92}.

We want to analyze the behaviour of iterations of a typical  infinite order
element of our groups $\G$ such as  the product of two generating involutions.
The first example we will examine is of great interest for statistical
mechanics on the lattice. It comes from the general 2-state vertex model on the
square lattice in dimension 2, i.e. the  sixteen-vertex model.
 Take $R$ to be  the general $2\times 2 $ matrix.
This matrix up to a multiplicative constant represents an element of the
15-dimensional projective space $\CP_{15}$.
A typical orbit of the adequate $\G$, when projected to a 2-plane of
coordinates,  looks like:
\par
\bigskip
\epsfxsize=\figwidth
\centerline { \epsffile[0 0 512 512]{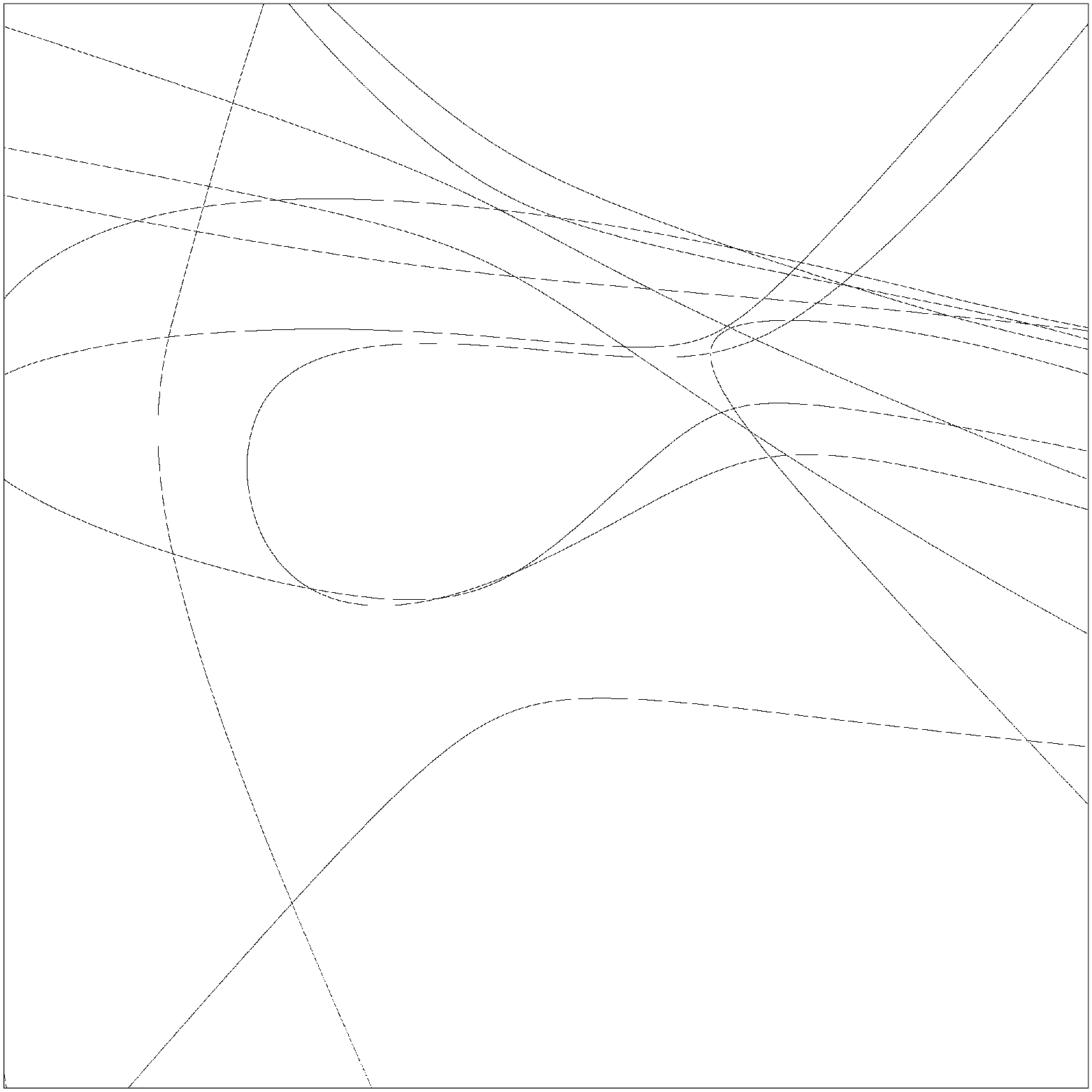} } \
\centerline{\it  Figure 2: One generic  orbit of $\Gamma$ for the 16-vertex
model}\par
\bigskip

This figure shows how efficient the  graphical method can be.
The situation is very favorable since the orbit is confined to  a low
dimensional variety (a curve).
What can be proved~\cite{BeMaVi92}, following the ideas of section
(\ref{invariants}), is the existence of a collection of invariants, of
algebraic rank $14$.
More precisely,  there are $17$ invariants with $3$ relations.
As in Figure 1, the discrete orbits are generically dense in a curve, and this
helps making the `graphical detector of invariants' so useful.

\section{More pictures} \label{morpic}

One should not believe that  $\G$-orbits are always curves.
Let us introduce a class of transformations in $\CP_2$ which contains in
particular the transformations (\ref{jaeger}).
Consider $\G$ generated by $I$ and $J$, given in homogeneous coordinates
$[x_0,x_1,x_2]$ by:
\begin{eqnarray} \label{coli}
& J: [x_0,x_1,x_2]  \rightarrow [x_1 x_2, x_0 x_2, x_1 x_2] & \label{colij}\\
& I = C^{-1} J C & \label{colii}
\end{eqnarray}
with $C$ a projective linear transformation of $\CP_2$.
This class of realizations, parametrized by the collineation matrix $C$ already
contains all kinds of behaviours. Moreover, many nice examples obtained from
matrix inversions do implement relation (\ref{colii}).

\subsection{Non-symmetric $Z_7$}

Suppose $m$ is the following cyclic  $7\times 7$ matrix:
\begin{equation}
m=\pmatrix{	x&	y& 	y&	z&	y&	z&	z\cr
		z&	x& 	y&	y&	z&	y&	z\cr
		z&	z& 	x&	y&	y&	z&	y\cr
		y&	z& 	z&	x&	y&	y&	z\cr
		z&	y& 	z&	z&	x&	y&	y\cr
		y&	z& 	y&	z&	z&	x&	y\cr
		y&	y& 	z&	y&	z&	z&	x\cr}
\end{equation}
The matrix inverse verifies (\ref{colii}) with\footnote{This form is a
particular value of an eigenmatrix of conference digraph~\cite{Ik93}}
\begin{equation} \label{cz7}
C_{Z_7}
=\left[\matrix{
        2       &6      &6      \cr
        2       &{{-1-i\sqrt{7}}}       &{{-1+i\sqrt{7}}} \cr
        2       &{{-1+i\sqrt{7}}}       &{{-1-i\sqrt{7}}}\cr
}\right]
\end{equation}

The iteration of $\varphi=IJ$, with $I$ the matrix inversion and $J$ the
element by element inversion, yields the following picture, in the
inhomogeneous coordinates $u=y/x$, $v=z/x$, if we represent a number of orbits
at the same time:
\par\bigskip
\epsfxsize=\figwidth
\centerline{ \epsffile[0 0 512 512]{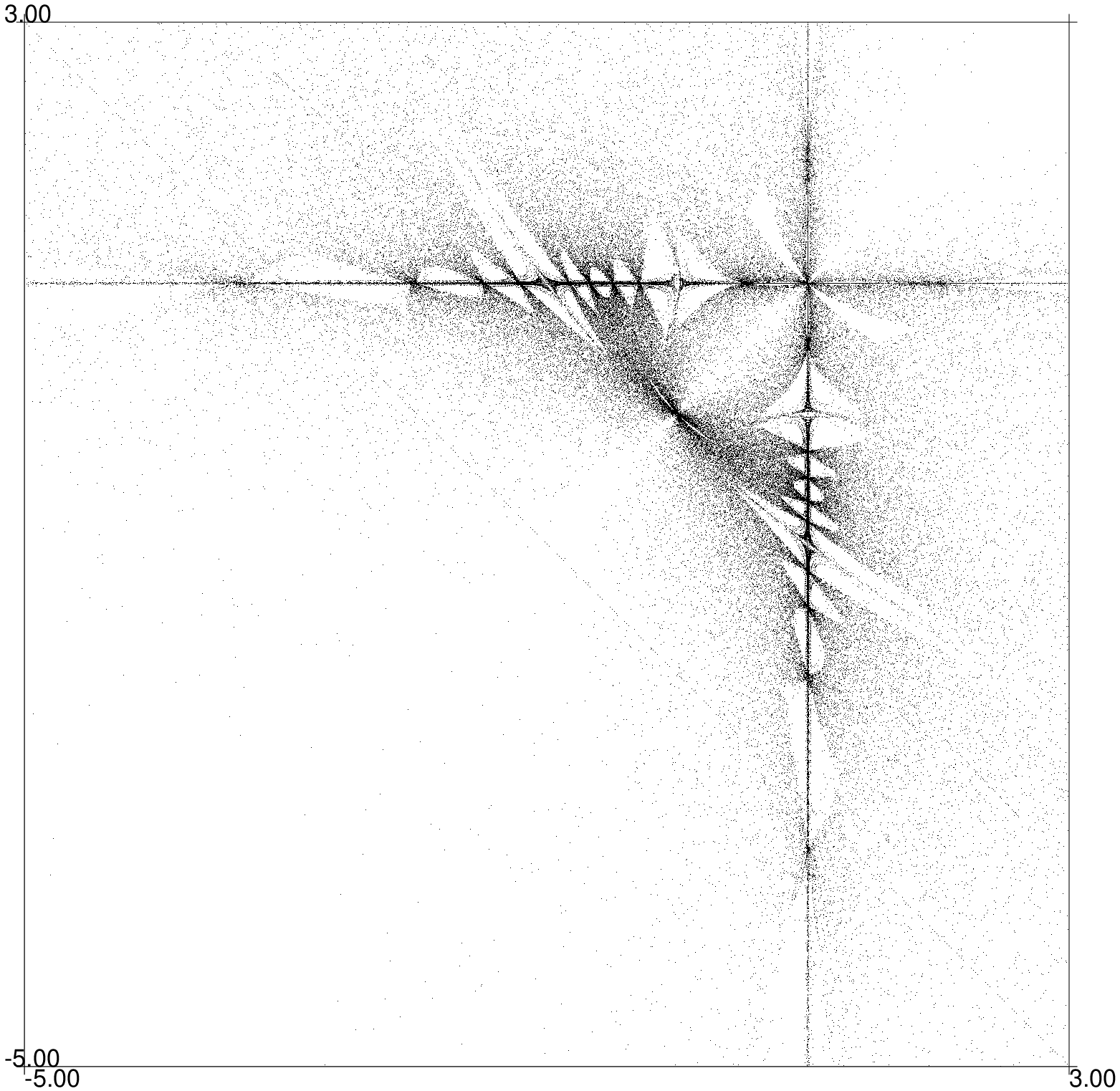} }
\centerline{\it  Figure 3: A  collection of chaotic orbits}\par
\smallskip

Figure 3  shows there should not be any invariant. Indeed, as proved
in~\cite{FaVi93}, there cannot be any because $\G$ has an infinite number of
singular points.
The first singular points are visible on the figure. Their location is very
easy to compute.

\subsection{A finite diagram model}

Take
\begin{equation} \label{cfdm}
 C = \left[\matrix{	2       &0      &2\cr
			1       &1      &-1\cr
		-1      &1      &1\cr } \right]
\end{equation}
This has been introduced in~\cite{FaVi93} and the denomination `finite diagram
model' refers to the finite number of singular points of $\G$.
Although such a transformation was not directly obtained from a matrix inverse,
it (or similar ones) appears in properly defined reductions of the inversion of
matrices of larger sizes~\cite{BoMaRo93b}.
We have found no invariant for this model. The graphical method is quite
ineffectual at generic points. It proves nevertheless  useful in the vicinity
of a fixed point.
\par\bigskip
\epsfxsize=\figwidth
\centerline { \epsffile[0 0 512 512]{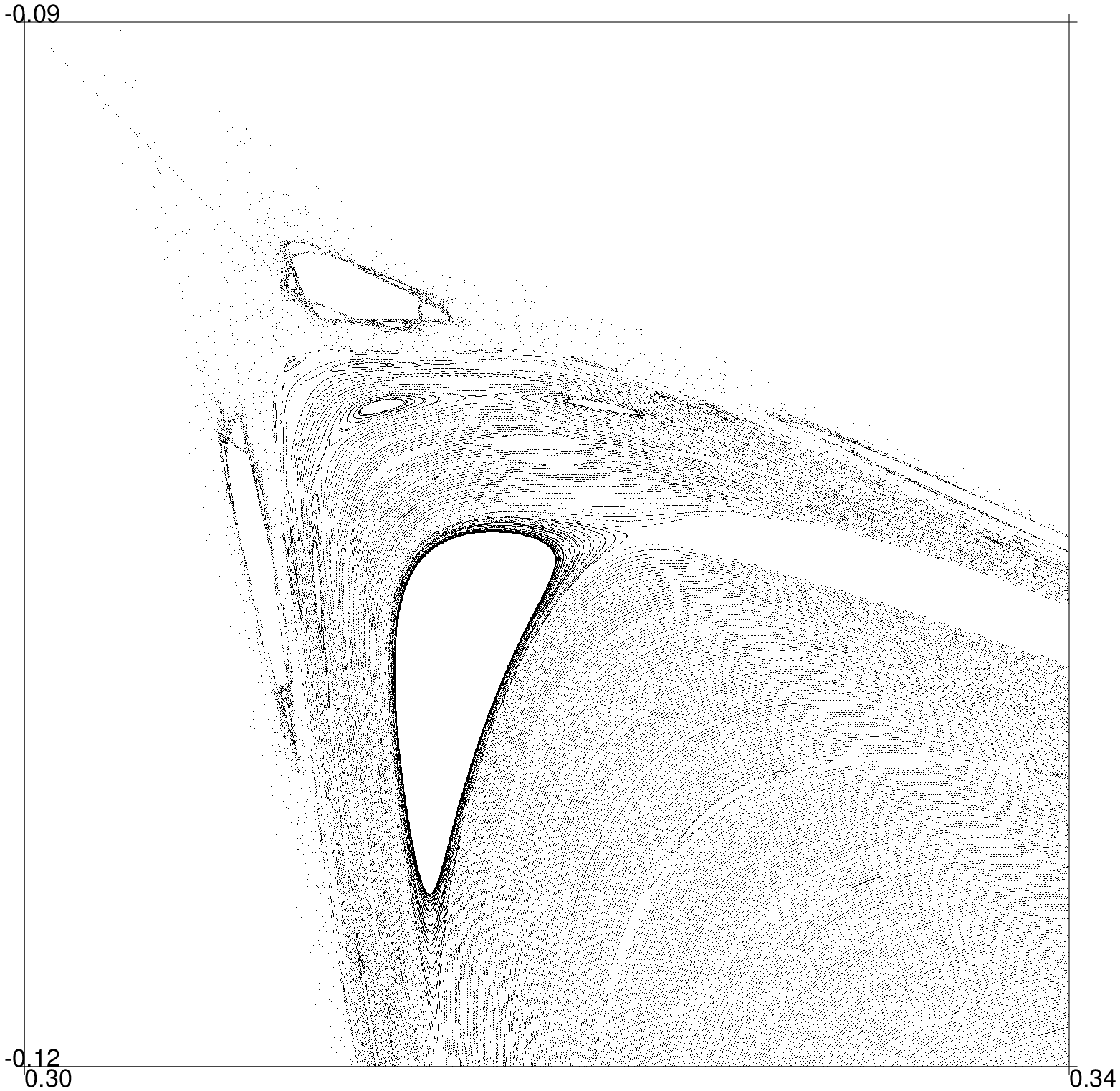} }
\centerline{\it  Figure 4}
\par
\bigskip
Figure 4  represents a collection of orbits not far away from a fixed point of
$(IJ)^2$. The fixed point $[\alpha^2,\alpha, -1], \alpha=(1+\sqrt{5})/2$,
$(x_1/x_0 \simeq .618, x_2/x_0 \simeq  -.382)$ is an elliptic fixed point at
which the eigenvalues of the linear differential are the roots of $s^2 -
{17/16} \; s+1$. We see that the orbits explode when one gets away from the
fixed point. This is exactly the situation encountered in~\cite{He69},
see~\cite{Gu91}.

\section{The complexity of iterations} \label{complexity}

More insight into the behaviour of the iterations can be gained by looking at
their complexity, in a way we will define.

The first grasp at the complexity of $\G$ is through its size (section
\ref{invariants}), but we shall not comment on this here.

We may satisfy ourselves with examining   the simple case $\nu=2$, where the
group is essentially two copies of $\bf Z$, and describe in the next section a
specific example (section \ref{classique}) with $\nu=3$.

The simplest situation  is the one where  $\G$ acts by transformations of the
2-dimensional projective space $\CP_2$.
We have given in section~\ref{morpic} a class  of  such a realization, and the
study of `the dynamics of singularities' as explored in~\cite{FaVi93} proves
extremely useful.

It is nevertheless something else we have in mind here:
suppose $\varphi=IJ$ is of degree $w$. Its iterates $\varphi^k$ do not have to
be of degree  $w^k$.
Indeed  there is a simple mechanism for the lowering of the degree $d(k)$  of
$\varphi^{k}$, for one has  to factorize out common factors from the
expressions of the homogeneous coordinates of  $\varphi^{(k)}$.
We may analyze the sequence $d(k)$, as a function of the order of the iteration
$k$.
This measure of the complexity of the iterations actually coincides with the
one introduced by Arnold~\cite{Ar90}; see~\cite{Ve92} for the particular case
of bi-polynomial transformations of the plane.
It brings  a variety of behaviours, lying between  the generic exponential
growth and periodicity, with the particular instance of polynomial (or
polynomially bounded) growth.
In a given system of coordinates the notion of degree is well defined. Of
course, changes of coordinates may change the degree, but allowed birational
changes will preserve the  nature of the growth.

We have calculated the degree of the successive iterates of $\varphi$ in a
number of cases, the simplest ones being given by the class
(\ref{colij},\ref{colii}).

If $C=C_{Z_7}$  given by  (\ref{cz7}), the first terms are:
\begin{eqnarray} \label{sequ7}
1, \qquad 4, \qquad  12, \qquad 33 ,\qquad 88, \qquad 232,  \qquad 609, \qquad
\dots
\end{eqnarray}
This may be seen as the first terms of the sequence:
\begin{eqnarray*}
d(k) = \sum_{i=0}^{k} f_{2 i}
\end{eqnarray*}
with $f_i$ the Fibonacci sequence:
\begin{eqnarray*}
1,\quad 2, \quad 3 ,\quad  5, \quad 8, \quad 13 ,\quad 21,  \quad 34, \quad
\dots
\end{eqnarray*}
This behaviour is found for { any {\em generic} elements of the four parameter
family}~\cite{Ne89} of collineations,
\begin{eqnarray} \label{neumaier}
\left [\begin {array}{ccc} 1&p&q-p-1\\1&s&-s-1\\1&r&-r-1\end {array}
\right ]
\end{eqnarray}
Notice that the Jaeger-Higman-Sims model belongs to this family (compare
collineations (\ref{jaeger}) and (\ref{neumaier})), and has the same sequence
(\ref{sequ7}).

For the `finite diagram model' with $C$ given by (\ref{cfdm}) the degree
reaches its maximal value:
\begin{eqnarray*}
d(k) = 4^k
\end{eqnarray*}

For \begin{equation} \label{cbmv}
C(w)=\left[\matrix{1&w-1&w\cr 1&-1&0\cr 1&0&-1\cr}\right],
\end{equation}
 $\G$ is known to have an invariant, and the orbits lie within elliptic curves
of equation
$$P_w^2(x_0,x_1,x_2) Q_w(x_0,x_1,x_2) - \lambda
(x_1+x_2)^4(x_0-x_2)^2(x_0-x_1)=0$$
where
$$P_w=(1-w)(x_2^2-x_0x_1)+(w-3)x_2(x_0-x_1),$$
$$
Q_w=(1-w^2)(x_1^3-x_0x_2^2)+(w^2-4w-1)x_1^2(x_0-x_2)+2(w-1)^2x_1x_2(x_0-x_1)$$
and the sequence of degrees starts with:
\begin{eqnarray*}
1 ,\;    4 ,\;   10 ,\;   20 ,\;   34  ,\;   53 ,\;   75 ,\;   102 ,\;   132
,\;   167 ,\;   206 ,\;   249 ,\;   295,\;   347,\;   402,\;   461,\;   \dots
\end{eqnarray*}
that is to say the second derivative $d(k+1)-d(k-1)-2d(k)$ is periodic of
period 12.
Notice that (\ref{cbmv}) has the form (\ref{neumaier}). It is just not generic.

As a last example, in  the Baxter model, and also for the 16-vertex model,  the
iteration of $\varphi=t_g I$ yields  the sequence of degrees:
\begin{eqnarray} \label{quadra}
1 , \quad  3 , \quad 9 , \quad 19 , \quad 33  , \quad 51 , \quad 73 , \quad 99
, \quad 129 , \quad 163 , \quad 201 , \quad  \dots
\end{eqnarray}
from which one infers
$$ d(k) = 2 k^2 +1 $$

We have examined many more examples, and they all  lead to the
conjecture~\cite{FaVi93}:
{\em  chaotic behaviour implies exponential growth, as regular behaviour
implies polynomial bounds}.

The proof of this conjecture should follow from considerations on addition on
elliptic curves~\cite{bbb6} at least for realizations in $\CP_2$, the main idea
being that $\varphi$ is a constant shift on some  curve.

Another approach goes through the study of the factorization properties: the
reduction of the degree $d(k)$ from  exponential to polynomially bounded  comes
from a sufficiently regular factorization process.
An analysis of this process was successfully undertaken
in~\cite{BoMaRo93a,BoMaRo93b} for various realizations $\G$ and confirms up to
now our conjecture. We shall give one more  example in the next section.

\section{An example}
\label{classique}

In~\cite{BeBoMaVi93} we introduced a restriction of the general 2-state model
on the cubic lattice in three dimensions, by imposing the following relations
on the entries of the $R$-matrix:
\begin{eqnarray}
\label{v1}
        R^{i_1 i_2 i_3}_{j_1 j_2 j_3}&=&R^{-i_1,-i_2,- i_3}_{-j_1,- j_2,-
j_3}\\\label{v2}
        R^{i_1 i_2 i_3}_{j_1 j_2 j_3}&=&0 \quad \mbox{ if }\quad
                {i_1 i_2 i_3}{j_1 j_2 j_3}=-1
\end{eqnarray}
 $R$ is the direct product of two times the same $4\times4$
submatrix~\cite{BeMaVi91e}.
It is further possible to impose that this $4 \times 4$ matrix is symmetric,
since the product $t=t_l t_m t_r$ acts as its transposition.
\begin{equation}
\label{v3}
         R^{i_1 i_2 i_3}_{j_1 j_2 j_3} =  R^{j_1 j_2 j_3}_{i_1 i_2 i_3}
\end{equation}

We  use  the following notations for the 10 homogeneous entries of this
$4\times4$ submatrix:
\begin{equation}
        \left( \begin{array}{cccc}
                                a   & d_1 & d_2 & d_3 \\
                                d_1 & b_1 & c_3 & c_2 \\
                                d_2 & c_3 & b_2 & c_1 \\
                                d_3 & c_2 & c_1 & b_3
        \end{array} \right).
\label{M1}
\end{equation}
The generators of $\G$ are the inversion of $R$, which acts as a matrix inverse
on (\ref{M1}), and the permutations of entries  $t_l$, $t_m$, and $t_r$ being
respectively:
\begin{eqnarray*}
& t_l:& \qquad c_3 \leftrightarrow d_3, \qquad c_2  \leftrightarrow d_2 \\
& t_m:&  \qquad c_1 \leftrightarrow d_1, \qquad c_3  \leftrightarrow d_3 \\
& t_r:&  \qquad c_2 \leftrightarrow d_2, \qquad c_1  \leftrightarrow d_1 \\
\end{eqnarray*}
A numerical iteration gives the following picture of an orbit, projected on a
2-plane of coordinates.
\par\bigskip
\epsfxsize=\figwidth
\centerline{ \epsffile[0 0 512 512]{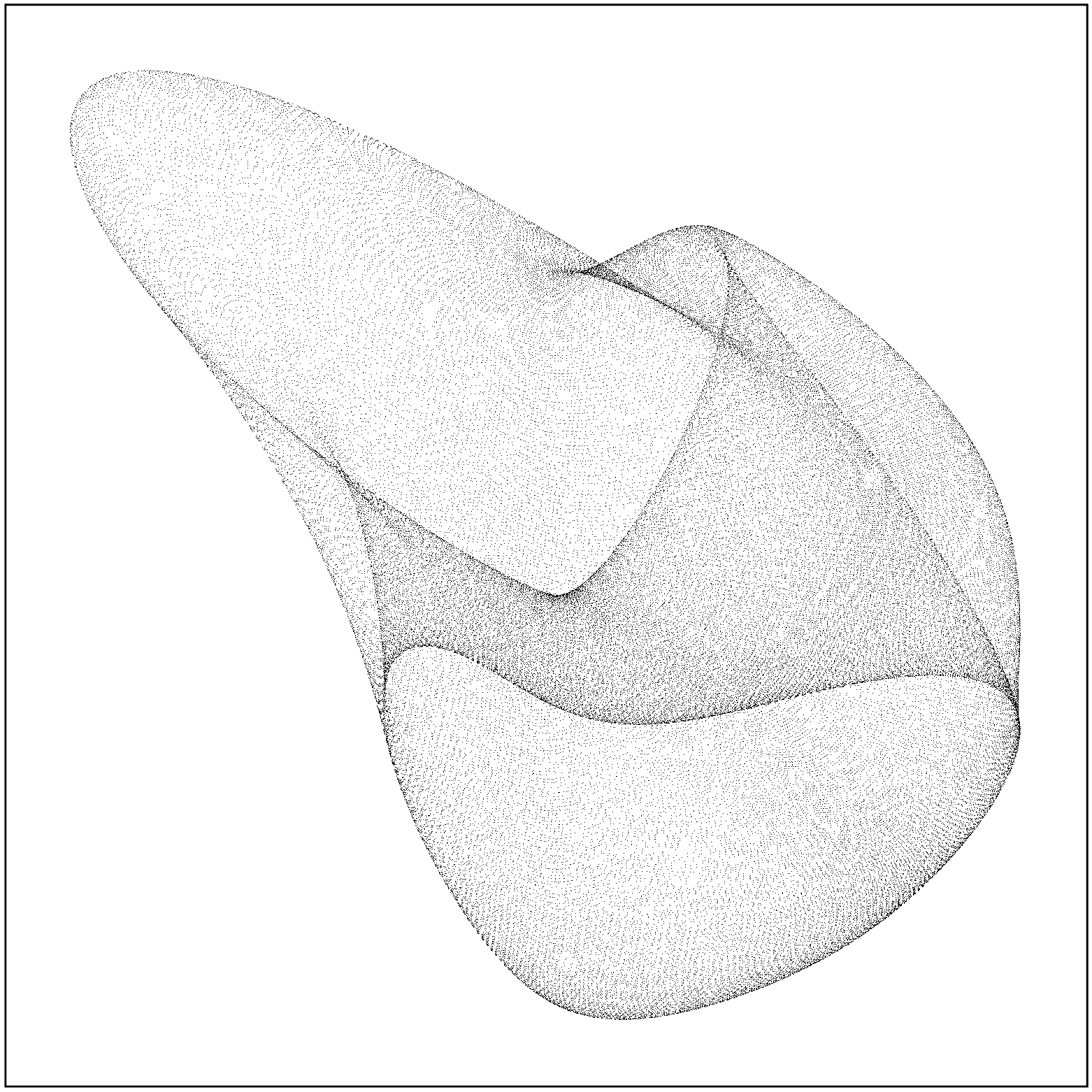} }
\centerline{\it  Figure 5: A `two-dimensional' orbit}\par
\bigskip
This is another example of the efficiency of the graphical detector: we see
that the orbit is confined to a 2-dimensional surface. This may be easily
proven.
If one defines
\begin{equation}
\label{pols}
        p_3 = a b_3 + b_1 b_2 - c_3^2 -d_3^2, \qquad
        q_3 = c_1 d_1 - c_2 d_2,
\end{equation}
together with the polynomial $p_2$ and $p_1$ obtained by permutation of the
indices 1,2, and 3, and (with a corrected misprint from~\cite{BeBoMaVi93}):
\begin{eqnarray*}
        r_3 &=& a b_3 - b_1 b_2 + c_3^2 - d_3^2,\\
        s_3 &=& (a+b_3) c_3 -d_1 d_2 -c_1 c_2, \\
        t_3 &=& (b_1 + b_2) d_3 - d_1 c_2 - c_1 d_2.
\label{more}
\end{eqnarray*}
then
\begin{itemize}
\item{ The polynomials $p_i, q_i, i=1,2,3$ form a five dimensional space, and
any ratio of these polynomials is invariant by the {\em whole group} $\G$.
This means that the orbit of $\G$ is confined to a 5-dimensional subspace of
the original 9-dimensional parameter space.}
\item{If we consider the subgroup generated by  $I$ and $t_r$, then
$r_3,s_3,t_3$ are three more covariant polynomials, and they furnish three more
invariants, proving that the orbit is confined to 2 dimensions.}
\end{itemize}

As far as the complexity is concerned, the non-chaotic image fits with the
sequence of degrees, which is the same as (\ref{quadra}).
The analysis of the degrees can be performed easily:
Let $M$ be a generic matrix of the form (\ref{M1}), and let $M_0=t_rM$ be the
starting point of the iteration. We iterate $t_rI$ defined polynomially in
terms of the entries, and keep  track  of the common factors (factorized out
from the  result of the action of $K$).
Let $M_{k+1}=t_rI(M_k)$, $f_k$ the extracted factor, and $\Delta_k=det(M_k)$.
One gets $d_0=det(t_rM)$, $f_0=1$, $d_1=det(t_rIt_rM) $, $f_2=1$.
Then a factorization of the determinants appears regularly, defining in the
course an additional sequence $\delta_k$ such that:
$$
\begin{tabular}{ccccccc}
 $ \Delta_2=\Delta_0^3\delta_2 $&$ \Delta_3=\Delta_1^3\delta_3 $&$
\Delta_4=\delta_2^3\delta_4$&$ \Delta_5=\delta_3^2\delta_5 $ 
&$\dots $&$\Delta_n=\delta_{n-2}^3\delta_n $\cr
$\dots$ & $f_3=\Delta_0^2 $&$ f_4=\Delta_1^2$&$
f_5=\Delta_2^2/\Delta_0^6=\delta_2^2 $ 
&$ \dots $&$ f_n=\delta_{n-3}^2 $
\end{tabular}
$$
Denoting by $u_k, v_k, g_k, x_k$ the degrees of respectively $\Delta_k,
\delta_k, f_k$ and of the matrix elements of $M_k$ in terms of the entries of
$M_0$, this implies
\begin{eqnarray*}
u_k& =& 3 \; v_{k-2} + v_k \\
g_k &=& 2 \; v_{k-3} \\
u_k &=& 4 \; x_k \\
x_k &=& 3 \; x_{k-1} - g_k
\end{eqnarray*}
resolved by
$$ x_{k+3} - 3 \;  x_{k+2} +3 \; x_{k+1} - x_k =0 \mbox{ and } d(k) = 2 k^2 +1
$$
This is an  instance of the general result obtained in~\cite{BoMaRo93b}
(formula (3.19), for $q=4$), giving a good flavour of the method.

\section{Conclusion and perspectives}

Inversion relations have proved to be  a powerful tool in the study of
statistical mechanical models, leading in particular to exact relations on the
partition functions, even for non-integrable or higher dimensional
models~\cite{Ba82,JaMa82,HaMaOiVe87}.

The description often appeals to a diagrammatic representation~\cite{Ba78},
which we have not used here.

The Yang-Baxter equations in their various forms have brought an enormous
amount of exact results in the field. They have unified branches of physics and
mathematics such as ice models~\cite{Li67,Li67b} and solitonic wave
equations~\cite{ZaFa71,FaTa86}, gave new insights into knot
polynomials~\cite{Ka87,Tu88,Jo89b,Wu92}, and produced successful offsprings
like quantum groups~\cite{Ji85,Dr86,Wor87}.

What~\cite{BeMaVi91c,BeMaVi91d} contain is a concrete link between discrete
dynamical systems and quantum integrability, producing a large number of
interesting dynamical systems.
We have been able to use a little part of the paraphernalia available from the
field of dynamical systems --the simplest being numerical calculations--.
One outcome is that the rational nature of the transformations  captures a good
proportion of the algebro-geometric content of the Yang-Baxter
equations~\cite{Kr81,BeMaVi92}.

One issue for future work is the resolution of the tetrahedron equations, and
the finding of truly three-dimensional integrability. Indeed  there is a
competition then between the higher overdetermination of the systems of
equations to solve, and the increasing size of the corresponding  group $\G$.
This antagonism might be resolved either by a relative triviality of the
solutions, which are disguised two-dimensional solutions, either by a low
complexity of the actual realization of $\G$, and this leaves room for
interesting solutions.
 The most natural to conjecture is that $\G$ has a {\em finite realizations},
or that its trajectories lie on abelian varieties~\cite{Ko93}.

The remarkable elliptic parametrization~\cite{BeMaVi92} of the 16-vertex model
should lead to interesting properties of the corresponding transfer matrices,
through controlled functional relations on the exact partition function  and/or
via a construction of physically relevant spaces of states. This applies as
well to three-dimensional models~\cite{BeBoMaVi93}, and prompts us to  a return
to basics, i.e: Bethe Ansatz~\cite{Be31}.
\par

\proclaim Acknowledgment. I would like to thank F.~Jaeger for a  very useful
discussion, and J.~Avan, M.~Bellon, J.-M.~Maillard, G.~Rollet, M.~Talon for
continued stimulating exchanges.\par

\end{document}